\documentstyle[aps,prl,epsfig,twocolumn]{revtex}

\begin{document}

\twocolumn[\hsize\textwidth\columnwidth\hsize\csname
@twocolumnfalse\endcsname
\title{Exact Insulating and Conducting Ground States of a Periodic Anderson
Model \\ in Three Dimensions}
\author{Zsolt~Gul\'acsi$^{a,b}$ and Dieter~Vollhardt$^{a}$}
\address{$^{(a)}$ Theoretical Physics III, Center for Electronic Correlations
and Magnetism,  Institute for Physics, University of Augsburg,
D-86135 Augsburg, Germany\\ $^{(b)}$ Department of Theoretical
Physics, University of Debrecen, H-4010 Debrecen, Hungary }
\date{February 10, 2003}
\maketitle
\begin{abstract}
We present a class of exact ground states of a three-dimensional
periodic Anderson model at 3/4 filling. Hopping and hybridization
of $d$ and $f$ electrons extend over the unit cell of a general
Bravais lattice. Employing novel composite operators combined with
55 matching conditions the Hamiltonian is cast into positive
semidefinite form. A product wave function in position space
allows one to identify stability regions of an insulating and a
conducting ground state. The metallic phase is a non-Fermi liquid
with one dispersing and one flat band.

\end{abstract}
\pacs{PACS No. 05.30.Fk, 67.40.Db, 71.10.-w, 71.10.Hf, 71.10.Pm} ]

The periodic Anderson model (PAM) is the basic microscopic model
for the investigation of heavy fermion and intermediate valence
systems, e.g., compounds containing elements with incompletely
filled $f$ shells such as Cerium or Uranium\cite{int1}. In its
simplest form the PAM describes strongly correlated,
dispersionless $f$ electrons which couple, via a local
hybridization, to non-interacting conduction ($d$) electrons
hopping between nearest neighbor sites. For real systems this is
certainly an oversimplification since there is experimental
evidence for (i) a weak, but finite, dispersion of the $f$
electrons (especially in Uranium compounds) \cite{keto}, (ii)
non-local contributions to the hybridization, and (iii) hopping of
the $d$ electrons beyond nearest neighbors\cite{egy}.

Recently, investigations into the origin of the dramatic volume
collapse at the $\alpha \rightarrow \gamma $ transition in Cerium
have drawn attention to the possibility of a Mott metal-insulator
transition in the PAM.\cite {int5,int5b,int5a} Although such a
transition is usually associated with the half-filled Hubbard
model with nearest neighbor hopping, there exist remarkable
similarities between the two models \cite{int4}, especially if
both the $d$ electron hopping and $d-f$ hybridization in the PAM
connect nearest neighbor sites.\cite{int4,int3,int6,int9} These
results show that the spatial range of the hopping and
hybridization in the PAM are very important, even on a qualitative
level.

In this situation exact results on the existence of insulating and
metallic phases in the PAM and their dependence on a general set
of hopping, hybridization and interaction parameters are
particularly desirable. Exact results for the PAM are very rare
since, in contrast to the Hubbard model, there does not even exist
an exact solution of the PAM in dimension $D=1$. On the other hand
it has been possible to construct exact ground states of the PAM
in certain regions of parameter space, namely for infinite
repulsion of the $f$ electrons\cite{a1,a2,a3}, and for finite
repulsion in low dimensions ($D=1,2$)\cite{a4,negy}.

In this Letter we show that it is possible to construct exact
ground state wave functions describing metallic and insulating
phases of the PAM at non-integer electron filling
{\em \ even in dimension} $D=3$. In particular, we explicitly
demonstrate (i) the insulating and conducting nature of the
solutions, (ii) the presence of strong variations in the
compressibility of the system when leaving the insulating phase,
and (iii) the non-Fermi liquid nature of the metallic phase.

We consider a general Bravais lattice in $D=3$ with a unit cell
$I$ defined by the primitive vectors $\{{\bf x}_{\tau }\}$, $\tau
=1,2,3$. The Hamiltonian of the PAM has the form
$\hat{H}=\hat{H}_{0}+U\hat{D}^{f}$ where $\hat{D} ^{f}=\sum_{{\bf
i}}\hat{n}_{{\bf i}\uparrow }^{f}\hat{n}_{{\bf i}\downarrow }^{f}$
describes the local Coulomb repulsion between the $f$ electrons
($U>0)$. The one-particle part may, in general, be written as
\begin{eqnarray}
&&\hat{H}_{0}=\sum_{{\bf i},\sigma }\biggm\{\sum_{{\bf r}}\left[
t_{{\bf r} }^{d}\hat{d}_{{\bf i}\sigma}^{\dagger }\hat{d}_{{\bf
i}+{\bf r},\sigma }+t_{{\bf r}}^{f}\hat{f}_{{\bf i}\sigma
}^{\dagger }\hat{f}_{{\bf i}+{\bf r} ,\sigma }+V_{{\bf
r}}(\hat{d}_{{\bf i}\sigma }^{\dagger }\hat{f}_{{\bf i}+ {\bf
r},\sigma }\right. \nonumber \\ &&\left. +\hat{f}_{{\bf i}\sigma
}^{\dagger }\hat{d}_{{\bf i}+{\bf r} ,\sigma })+H.c.\right]
+V_{0}(\hat{d}_{{\bf i}\sigma }^{\dagger }\hat{f}_{ {\bf i}\sigma
}+H.c.)+E_{f}\hat{n}_{{\bf i}\sigma }^{f}\biggm\},
\label{Hamstart}
\end{eqnarray}%
where the terms with $t_{{\bf r}}^{d,f}$ represent the kinetic
energy of $d$ and $f$ electrons due to hopping between two sites
${\bf i}$ and ${\bf i}+{\bf r}$, $V_{{\bf r}}$ is the
hybridization of $d$ and $f$ electrons at sites ${\bf i}$ and
${\bf i}+{\bf r}$, $ V_{0}$ is the local hybridization, and
$E_{f}$ is the local $f$ electron energy; here ${\bf r}\neq 0$.
The amplitudes $t_{{\bf r}}^{d,f}$ are real, but $V_{{\bf r}}$,
$V_{0}$ can, in principle, be complex. In our investigation the
hopping and hybridization of the $d$ and $f$ electrons extend over
the unit cell of a general Bravais lattice (Fig. 1). To avoid
multiple counting of contributions to (1) by the $H.c.$ term, the
vector ${\bf r}$ must be properly defined. To this end the sites
within $I_{{\bf i}}$, the unit cell defined at site ${\bf i}$, are
denoted by ${\bf r}_{I_{{\bf i} }}={\bf i}+{\bf r}_{\alpha \beta
\gamma }$, with ${\bf r}_{\alpha \beta \gamma }=\alpha {\bf
x}_{1}+\beta {\bf x}_{2}+\gamma {\bf x}_{3}$; $\alpha ,\beta
,\gamma =0,1$. As shown in Fig. 1 the eight sites ${\bf
r}_{I_{{\bf i} }}$ can be numbered by the indices $n(\alpha ,\beta
,\gamma )=1+\alpha +3\beta +4\gamma -2\alpha \beta $ without
reference to $I_{{\bf i}}$. In the following we use orthogonal
${\bf x}_{\tau }$ vectors for simplicity. Then $ {\bf r}={\bf
r}_{\alpha ^{\prime }\beta ^{\prime }\gamma ^{\prime }}-{\bf r}
_{\alpha \beta \gamma }$, with $n(\alpha \prime ,\beta \prime
,\gamma \prime )>n(\alpha ,\beta ,\gamma )$, connects any two
sites within a unit cell \cite {expl0}.

We now introduce a superposition of operators creating $b$ (= $d$
or $f$) electrons inside every unit cell $I_{{\bf i}}$ as

\begin{figure}[h]
\epsfxsize=6cm \centerline{\epsfbox{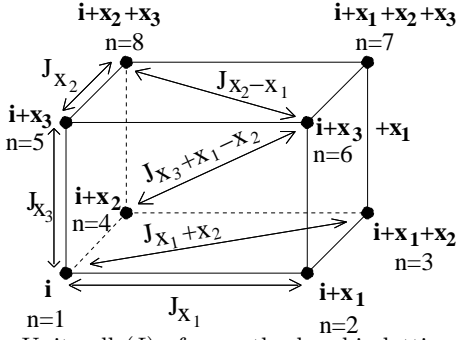}} \caption{Unit
cell ($I$) of an orthorhombic lattice at an arbitrary site ${\bf
i}$ showing the primitive vectors ${\bf x}_{\protect\tau}$ and
indices $n$ of the sites in $I$. Arrows depict some of the hopping
and hybridization matrix elements ($J=t,V$) extending over $I$.}
\label{fig1}
\end{figure}

\begin{eqnarray}
&&\hat{A}_{I_{{\bf i}}\sigma }^{\dagger }
=\sum_{b=d,f}\sum_{\alpha ,\beta ,\gamma =0}^{1}a_{n(\alpha ,\beta
,\gamma ),b}^{\ast }\hat{b}_{{\bf i}+{\bf r }_{\alpha \beta \gamma
},\sigma }^{\dagger } = \nonumber\\ &&\sum_{b=d,f}(a_{1,b}^{\ast
}\hat{b}_{{\bf i}\sigma }^{\dagger }+a_{2,b}^{\ast }\hat{b}_{{\bf
i}+{\bf x}_{1},\sigma }^{\dagger }+...+a_{8,b}^{\ast
}\hat{b}_{{\bf i}+{\bf x}_{2}+{\bf x}_{3},\sigma }^{\dagger }),
\label{aaa}
\end{eqnarray}
with $a_{n,b}^{\ast}\neq 0$ for all $n$. Although
$\{\hat{A}_{I\sigma }^{\dagger },\hat{A}_{I^{\prime }\sigma
^{\prime }}^{\dagger }\}=\{\hat{A}_{I\sigma },\hat{A}_{I^{\prime
}\sigma ^{\prime }}\}=0$ and $\{\hat{A}_{I\sigma
},\hat{A}_{I\sigma }^{\dagger }\}$=$ K_{d}+K_{f}$, where
$K_{b}=\sum_{n=1}^{8}|a_{n,b}|^{2}$ and $n\equiv n(\alpha ,\beta
,\gamma )$, the composite operator $\hat{A}_{I\sigma }^{\dagger }$
does not obey canonical anti-commutation rules since $\{\hat{A}
_{I\sigma },\hat{A}_{I^{\prime }\sigma ^{\prime }}^{\dagger
}\}\neq 0$ for $ I\neq I^{\prime }$. Due to the translational
symmetry of the lattice the prefactors $a_{n,b}^{\ast}$ {\em are
the same} in every unit cell. Making use of this fact, $\hat{H}$
can be cast into the form
\begin{eqnarray}
\hat{H}=\sum_{{\bf i},\sigma }\hat{A}_{I_{{\bf i}}\sigma
}\hat{A}_{I_{{\bf i} }\sigma }^{\dagger }+U\sum_{{\bf
i}}\hat{P}_{{\bf i}}+E_{g}, \label{HA}
\end{eqnarray}
where $\hat{P}_{{\bf i}}=\hat{n}_{{\bf i}\uparrow
}^{f}\hat{n}_{{\bf i} \downarrow }^{f}-\hat{n}_{{\bf i}\uparrow
}^{f}-\hat{n}_{{\bf i}\downarrow }^{f}+1$,
$E_{g}=K_{d}N+UN_{\Lambda }-2N_{\Lambda }(2K_{d}-E_{f})$, and $N$
and $N_{\Lambda }$ is the number of electrons and lattice sites,
respectively. For (3) to reproduce (1) the prefactors
$a_{n,b}^{\ast }$ in $ \hat{A}_{I_{{\bf i}}\sigma }^{\dagger }$
must be expressed in terms of the microscopic parameters $t_{{\bf
r}}^{d}$, $t_{{\bf r}}^{f}$, $V_{{\bf r}}$, $V_{{\bf r}}^{\ast }$,
$V_{0}$, $V_{0}^{\ast }$, $E_{f}$, $U$, for all ${\bf r}\in
I_{{\bf i}}$, taking into account periodic boundary conditions.
This leads to 55 coupled, non-linear matching conditions
\cite{condi} which can be written in compact notation, with
$b,b^{\prime }=d,f$, as \cite{55}
\begin{eqnarray}
\sum_{\beta _{1},\beta_{2},\beta_{3} = -1}^{1}
(\prod_{i=1}^{3}D_{\beta _{i},\alpha _{i}}) a_{n^{+},b}^{\ast
}a_{n^{-},b^{\prime }}=T_{\bar {\bf r},\nu }^{b,b^{\prime }}.
\label{cond}
\end{eqnarray}

Apart from the constant term $E_{g}$ in (\ref{HA}) $\hat{H}$ is
positive semidefinite operator. A state $|\Psi _{g}\rangle $
fulfilling the conditions $\hat{P}_{{\bf i}}|\Psi _{g}\rangle =0$
and $\hat{A}_{I_{{\bf i} }\sigma }^{\dagger }|\Psi _{g}\rangle =0$
for all ${\bf i}$ will then be an {\em exact ground state} of
$\hat{H}$ with energy $E_{g}$. We note that $\hat{P}_{ {\bf i}}$
assumes its lowest eigenvalue, $0$, when there is at least one $f$
electron on site ${\bf i}$. Therefore a state of the form $|\Psi
_{g}\rangle \sim \prod_{{\bf i}=1}^{N_{\Lambda }}\hat{F}_{{\bf
i}}^{\dagger }|0\rangle $ , where the operator $\hat{F}_{{\bf
i}}^{\dagger }=\mu _{{\bf i}\uparrow } \hat{f}_{{\bf i}\uparrow
}^{\dagger }+\mu _{{\bf i}\downarrow }\hat{f}_{{\bf i}\downarrow
}^{\dagger }$ (with arbitrary coefficients $\mu _{{\bf i}\sigma
}$) creates one $f$ electron on every site ${\bf i}$, fulfills the
first condition. Further, a state of the form $|\Psi _{g}\rangle
\sim \prod_{ {\bf i}=1}^{N_{\Lambda }}(\hat{A}_{I_{{\bf
i}}\uparrow }^{\dagger }\hat{A} _{I_{{\bf i}}\downarrow }^{\dagger
})|0\rangle $ fulfills the second condition. Consequently, the
(unnormalized) product state
\begin{eqnarray}
|\Psi _{g}\rangle =\prod_{{\bf i}=1}^{N_{\Lambda }}\left[
\hat{A}_{I_{{\bf i} }\uparrow }^{\dagger }\hat{A}_{I_{{\bf
i}}\downarrow }^{\dagger }\hat{F}_{ {\bf i}}^{\dagger }\right]
|0\rangle \label{psi0}
\end{eqnarray}
has at least one $f$ electron on every site ${\bf i}$, the product
$\prod_{ {\bf i}=1}^{N_{\Lambda }}\hat{A}_{I_{{\bf i}}\uparrow
}^{\dagger }\hat{A} _{I_{{\bf i}}\downarrow }^{\dagger }$ creating
at most two more ($d$ or $f$ ) electrons on ${\bf i}$. Clearly,
$|\Psi _{g}\rangle $ has the desired property $\hat{H}|\Psi
_{g}\rangle =E_{g}|\Psi _{g}\rangle $ and is thus an exact ground
state of $\hat{H}$ with energy $E_{g}$. Since the product of the
three operators in $|\Psi _{g}\rangle $ creates $N=3N_{\Lambda }$
electrons, the ground state is $3/4$ filled, i.e., there are on
average three electrons per site. The arbitrariness of $\mu _{{\bf
i} ,\sigma }$ implies a large spin degeneracy of $|\Psi
_{g}\rangle $ which is globally paramagnetic. Neglecting the
trivial $2S+1$ multiplicity of the { spin} orientation, the ground
state is $N_{\Lambda }/2$-fold degenerate\cite {condi3}; all
degeneracies are still contained in (5). We note that $|\Psi
_{g}\rangle $ is only a ground state of the {\em interacting}
system since the operator $\hat{F}_{{\bf i}}^{\dagger }$ in $|\Psi
_{g}\rangle $ is required to enforce $U\sum_{{\bf i}}\hat{P}_{{\bf
i}}|\Psi _{g}\rangle =0$ for $U>0$. Hence $|\Psi _{g}\rangle $ is
not connected in any simple (e.g., perturbative) way to the
non-interacting ($U=0 $) ground state of the model.

The physical nature of $|\Psi _{g}\rangle $ depends on the values
of the coefficients $a_{n,b}^{\ast }$ in (\ref{aaa}) which are
solutions of (\ref{cond}) for given microscopic parameters. We
will now identify localized and itinerant ground states, and
discuss their physical properties.

{\it 1. Localized ground state.} The coefficients $a_{n,b}$ may be
chosen in such a way that $|\Psi _{g}\rangle $ has {\em exactly}
three electrons per site. To see how this can be achieved we take
a look at a typical factor $\hat{A} _{I_{{\bf j}}\uparrow
}^{\dagger }\hat{A}_{I_{{\bf j}}\downarrow }^{\dagger }
\hat{A}_{I_{{\bf j}^{\prime }}\uparrow }^{\dagger
}\hat{A}_{I_{{\bf j} ^{\prime }}\downarrow }^{\dagger
}\hat{F}_{{\bf j}^{\prime \prime }}^{\dagger }$ entering in $|\Psi
_{g}\rangle $, where ${\bf j}^{\prime \prime }$ is a common site
of the two unit cells $I_{\bf j}$, $ I_{{\bf j}^{\prime }}$. Since
$\hat{F}_{{\bf j}^{\prime \prime }}^{\dagger }$ always creates one
$f$ electron on ${\bf j}^{\prime \prime }${\bf ,} the product of
the four unit cell operators should create only two more electrons
(one $ \uparrow $ and one $\downarrow $) on ${\bf j}^{\prime
\prime }$. Therefore terms of the form $\hat{d}_{{\bf j}^{\prime
\prime }\uparrow }^{\dagger }\hat{f}_{{\bf j}^{\prime \prime
}\uparrow }^{\dagger }\hat{d}_{ {\bf j}^{\prime \prime }\downarrow
}^{\dagger }\hat{f}_{{\bf j}^{\prime \prime }\downarrow }^{\dagger
}(a_{n_{1},d}^{\ast }a_{n_{2},f}^{\ast }-a_{n_{1},f}^{\ast
}a_{n_{2},d}^{\ast })^{2}$ which are also generated and which lead
to more than two additional electrons on this site must be
prohibited. This can be achieved by choosing $a_{n,d}^{\ast
}/a_{n,f}^{\ast }\equiv p$ for all $n$. It follows from
(\ref{cond}) that along the diagonal of a unit cell with end
points $ n,n^{\prime }$  the relation $a_{n,f}^{\ast }a_{n^{\prime
},d}=a_{n,d}^{\ast }a_{n^{\prime },f}$ holds, implying $p$ and
therefore also the hybridization amplitudes to be real. A solution
of (\ref{cond}) of this form is obtained for $t_{{\bf
r}}^{b}=t_{\nu }^{b}$, i.e., for equal hopping along $x,y,z$. For
the coefficients chosen in this way the ground state becomes
\begin{equation}
|\Psi _{loc}\rangle =\prod_{{\bf i}} [ \sum_{\sigma }\mu _{{\bf
i}\sigma }p(p\hat{d}_{{\bf i}\downarrow }^{\dagger }\hat{d}_{{\bf
i}\uparrow }^{\dagger }\hat{f}_{{\bf i}\sigma }^{\dagger
}+\hat{f}_{{\bf i}\uparrow }^{\dagger }\hat{f}_{{\bf i}\downarrow
}^{\dagger }\hat{d}_{{\bf i}\sigma }^{\dagger }) ] |0\rangle .
\label{ploc}
\end{equation}
Denoting ground state expectation values by $\langle ...\rangle $
the local $ f,d$-electron occupations are found as $\langle
\hat{n}_{{\bf i}}^{f}\rangle =(1+2z)/(1+z),\langle \hat{n}_{{\bf
i}}^{d}\rangle =(2+z)/(1+z)$, where $ z= |t_{1}^{f}/t_{1}^{d}| $
is a measure of the nearest-neighbor hopping amplitude of the $f$
electrons. Since the latter can be expected to be much smaller
than that of the $d$ electrons ($z<<1$), the $f$ ($d$) electron
occupation per site is found to be close to one (two). Hence there
exist local moments on most of the $f$ sites. We see that $\langle
\hat{n}_{{\bf i}}\rangle =\langle \hat{n}_{{\bf i}}^{f}\rangle
+\langle \hat{ n}_{{\bf i}}^{d}\rangle =3$ for all ${\bf i}$,
i.e., the electron distribution is indeed uniform. The
localization is due to a subtle quantum mechanical interference
between the hopping and hybridization processes of the electrons.
Therefore the nature of this localized state is quite non-trivial.
\begin{figure}[h]
\epsfxsize=6cm \centerline{\epsfbox{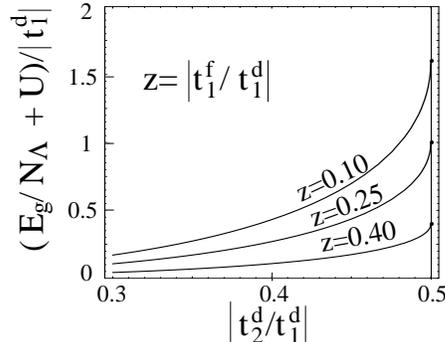}}
\caption{Ground-state energy of the localized state expressed in
terms of $( E_{g}/N_{\Lambda} + U)/\left| t_{1}^{d}\right|$ as a
function of next-nearest neighbor hopping of $d$ electrons,
$|t_{2}^{d}/t_{1}^{d}|$, for different values of $f$ hopping
$z=\left| t_{1}^{f}/t_{1}^{d} \right|$; here $\left|
V_{1}/V_{0}\right|
>1/2$.} \label{fig2}
\end{figure}
By separating the Hamiltonian $\hat{H}$ into an itinerant part
$\hat{H} _{itin}\equiv \hat{H}_{0}-\sum_{{\bf i},\sigma
}[V_{0}(\hat{d}_{{\bf i} \sigma }^{\dagger }\hat{f}_{{\bf i}\sigma
}+H.c.)+E_{f}\hat{n}_{{\bf i} \sigma }^{f}]\equiv \sum_{{\bf
r}}\hat{H}_{itin}({\bf r})$ and a complementary localized part
$\hat{H}_{loc} \equiv \hat{H}-\hat{H}_{itin}$, and using $\langle
\hat{b}_{{\bf i}\sigma }^{\dagger }\hat{b}_{{\bf j}\sigma ^{\prime
}}^{\prime }\rangle =0$ for ${\bf i}\neq {\bf j}$ and $b,b^{\prime
}=d,f$, one finds $\langle \hat{H}_{itin}({\bf r})\rangle =0$ and
$\langle \hat{H} _{loc}\rangle =E_{g}.$ This clearly illustrates
the localized, and thus insulating, nature of the ground state
whose energy is obtained as $ E_{g}=N_{\Lambda }[(1-2z)K_{f}/z-U]$
(Fig. 2). The parameter space in which the localized ground state
$|\Psi _{loc}\rangle $ is stable is represented in terms of the
variables $ E_{f},U,t_{1}^{f}/t_{1}^{d},t_{2}^{d}/t_{1}^{d}$ by
the surfaces $I_1$, $I_2$ in Fig. 3. The stability region is seen
to extend through the phase diagram from weak to strong
interactions $U$.

For $y=\left| t_{2}^{d}/t_{1}^{d}\right| >1/2$, i.e., rather large
next-nearest neighbor hopping of $d$ electrons, the localized
state $ |\Psi _{loc}\rangle $ ceases to be the ground state.
Apparently, at $ y_{c}=1/2$ a different, most probably itinerant,
phase becomes stable. We note that the ground state energy
$E_{g}(y)$ has a finite value at $y_{c},$ but {\em infinite} slope
(see Fig.2). This exact result has a direct physical
interpretation. Namely, since the size of the hopping element may
be tuned by pressure, the infinite slope of $E_{g}$ at $y=y_{c}$
is expected to correspond to an anomaly in the compressibility at
a critical pressure $P_{c}$. Such a feature is indeed observed in
some heavy-fermion systems \cite{het}.

{\it 2. Itinerant ground state}. The localized ground state
discussed above has exactly three electrons per site. In general,
the intersite hopping and hybridization will lead to a variable
number of electrons per site. For such states $\langle
\hat{H}_{itin}\rangle \neq 0$, implying a non-zero conductivity of
the system as expressed by the sum-rule $\int_{0}^{\infty }d\omega
Re\sigma _{\tau ,\tau }(\omega )=-\pi e^{2}|{\bf x}_{\tau
}|^{2}/(2\hbar V)\sum_{{\bf r},{\bf r}\cdot {\bf x}_{\tau }\neq
0}\langle \hat{H}_{itin}({\bf r})\rangle ,$ where $\sigma _{\tau
,\tau }(\omega )$ is defined by the Kubo formula \cite{cur}. An
itinerant ground state is obtained, for example, by choosing
$p_{n}^{\ast }=-p_{n},|p_{n}|=|p|$, corresponding to imaginary $p$
and, hence, imaginary hybridization amplitudes $\ V_{{\bf r}}$.
Whether $V_{{\bf r}}$\ is real, complex or imaginary depends on
the linear combination of the corresponding electronic orbitals
\cite{int4,negy,nyo,kil,sham}, and hence on the lattice symmetry.
For example, $V_{{\bf r}}$ may be tuned from real to imaginary by
introducing axial distortions of $D_{4h}$ symmetry to an
underlying $O_{h}$ lattice symmetry \cite{kil}. Therefore such a
solution requires anisotropic hopping and hybridization
amplitudes. The itinerant ground state discussed here emerges if
the hybridization on the same site and in the basal ($xy$) plane
vanish. The anisotropy in the hopping starts at the level of
next-nearest neighbor amplitudes.

To describe the itinerant case a ${\bf k}$-type representation is
more
suitable. 
Denoting the Fourier transforms of $\hat{A}_{I_{{\bf i}}\sigma }$
and $\hat{b}_{{\bf i}\sigma }$ by $\hat{A}_{{\bf k}\sigma }$ and
$\hat{b}_{{\bf k}\sigma }$, respectively, (\ref{aaa}) takes the
form $\hat{A}_{{\bf k}\sigma }^{\dagger }=\sum_{b=d,f}a_{{\bf
k}b}^{\ast } \hat{b}_{{\bf k}\sigma }^{\dagger }$; the expressions
for the coefficients $a_{{\bf k}b}^{\ast }$ will not be reproduced
here. We can now define new canonical Fermi operators
$\hat{C}_{\delta ,{\bf k}\sigma }$, $\delta =1,2$, where $
\hat{C}_{1,{\bf k}\sigma }=R_{{\bf k}}^{1/2}\hat{A}_{{\bf k}
\sigma }$, with $R_{{\bf k}}^{-1}=\sum_{b}|a_{{\bf k}b}|^{2}$, and
$\hat{C}_{2,{\bf k}\sigma }$ is determined by the anti-commutation
rules between $\hat{C}_{1,{\bf k}\sigma }$ and $\hat{C}_{2,{\bf
k}\sigma }^{\dagger}$. It follows that $-\sum_{{\bf i},\sigma
}\hat{A}_{I_{{\bf i}}\sigma }^{\dagger } \hat{A}_{I_{{\bf
i}}\sigma }+K_{d}\hat{N}=\sum_{{\bf k},\sigma }\left[
(K_{d}-R_{{\bf k}}^{-1})\hat{C}_{1,{\bf k}\sigma }^{\dagger
}\hat{C}_{1,{\bf k} \sigma }+K_{d}\hat{C}_{2,{\bf k}\sigma
}^{\dagger }\hat{C}_{2,{\bf k}\sigma }
\right] \equiv \hat{H}_{g}$ 
, such that (\ref{HA}) can be written as
$\hat{H}=\hat{H}_{g}+U\hat{P}+N_{\Lambda}[U-2(K_{d}-E_{f})]$. In
the ground state, using $\hat{P}|\Psi _{g}\rangle =0 $, the
Hamiltonian $\hat{H}$ therefore reduces to $\hat{H}_{g}$. Thus we
succeeded in diagonalizing $\hat{H}$ for the ground state. %
There are two bands, the lower one having a dispersion
$K_{d}-R_{{\bf k}}^{-1}$, while the upper one is dispersionless
("flat"); the Fermi energy is $ E_{F}=K_{d}$. Such a
band-structure around $E_{F}$ has been observed in experiment
\cite{exp}.

The momentum distribution of the $d,f$ electrons becomes
\begin{eqnarray}
n_{{\bf k}}^{b}=\langle \sum_{\sigma }\hat{b}_{{\bf k}\sigma
}^{\dagger } \hat{b}_{{\bf k}\sigma }\rangle =(2|a_{{\bf
k}b}|^{2}+|a_{{\bf k,}b^{\prime }\neq b}|^{2})R_{{\bf k}} ,
\label{iti}
\end{eqnarray}
with $n_{{\bf k}}=\sum_{b=d,f}n_{{\bf k}}^{b}=3$. Since the
coefficients $a_{ {\bf k}b}$ are regular functions of ${\bf k}$
this also holds for $n_{{\bf k}}^{d}$, $n_{{\bf k}}^{f}$, and
$n_{{\bf k}}$. Consequently, the momentum distributions of the
electrons in the interacting ground state has no discontinuities.
Since the ground state is paramagnetic and metallic, the system is
a {\em non-Fermi liquid}. This is a consequence of the degeneracy
of electrons in the upper band. In terms of the $\hat{C} _{\delta
,{\bf k}\sigma }$ fermions one finds $\langle \hat{C}_{1,{\bf k}
\sigma }^{\dagger }\hat{C}_{1,{\bf k}\sigma }\rangle =1$ and
$\langle \hat{C} _{2,{\bf k}\sigma }^{\dagger }\hat{C}_{2,{\bf
k}\sigma }\rangle =1/2$ (upper band half filled).

The itinerant solution can be generalized to fillings beyond $3/4$
by inserting the operator
$\hat{V}_{M}^{\dagger}=\prod_{j=1}^{M}[\sum_{i=1}^{N_{\Lambda
}}a_{ji}(\sum_{b=d,f;\sigma }\epsilon _{b\sigma }\hat{b}_{i\sigma
}^{\dagger })]$ into (\ref{psi0}) next to $|0\rangle $; here
$a_{ji}$, $\epsilon _{b\sigma }$ are numerical coefficients. This
operator introduces $M<N_{\Lambda}$ additional particles into the
ground state. It allows one to calculated the energy $E_{g}$ for
different particle numbers. In particular, one finds $\mu
^{+}\equiv E_{g}(N+2)-E_{g}(N+1)=K_{d}$, $\mu ^{-}\equiv
E_{g}(N+1)-E_{g}(N)=K_{d}$, i.e., $\mu ^{+}-\mu ^{-}=0$. Therefore
the system is a conductor. \cite{geb} The stability region of the
conducting state corresponds to the surface $C$ in Fig. 3.

\begin{figure}[h]
\epsfxsize=6cm \centerline{\epsfbox{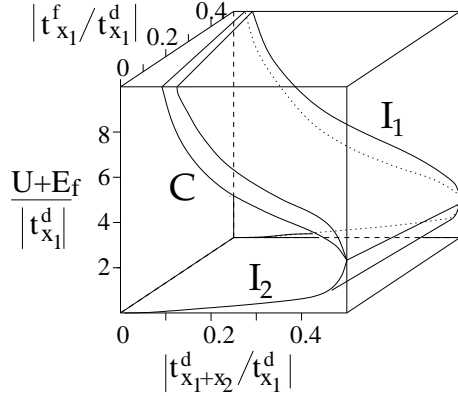}}
\caption{Surfaces in parameter space representing stability
regions of the insulating ground-state ($I_1$ for $\left|
V_{1}/V_{0}\right| <1/2$, $I_2$ for $\left| V_{1}/V_{0}\right|
>1/2$),  and the conducting ground-state ($C$).
} \label{fig3}
\end{figure}

In summary, we derived the first exact ground state solution of a
{\em three}-dimensional periodic Anderson model with finite
hopping and hybridization of $d$ and $f$ electrons in the unit
cell. This was achieved by (i) casting the Hamiltonian into a
positive semidefinite form using composite operators in
combination with 55 coupled, non-linear matching conditions, and
(ii) constructing a product wave function of these composite
operators in position space. For real hybridization amplitudes we
obtained an {\em insulating} ground state whose compressibility
diverges at the boundary of the stability region. By contrast, for
imaginary hybridization amplitudes we identified a {\em conducting
non-Fermi liquid} state consisting of one dispersing band and one
(upper) flat band. The stability regions of the two ground states
extend through an unexpectedly large region of parameter space,
e.g., from weak to strong interactions $U$.

By modifying the structure of the composite operators it is
possible to vary the stability regions of these ground states, and
also to describe magnetically ordered phases.

The authors thank R. Bulla, F. Gebhard, P. Gurin, K. Held, A.
Kampf, S. Kehrein, M. Kollar, Th. Pruschke, and M. Vojta for
valuable discussions. One of us (ZG) acknowledges support by the
Alexander von Humboldt foundation. This work was supported in part
by the Hungarian Scientific Research Fund (contract OTKA-T037212)
and the Deutsche Forschungsgemeinschaft through SFB 484.


\begin{references}
\bibitem{int1} P. A. Lee {\em et al.}, Comm.Cond.Matt.Phys. {\bf 12}, 99
(1986).

\bibitem{keto} A. J. Arko {\em et al.}, J. Elec. Spec. {\bf 117-118}, 323
(2001).

\bibitem{egy} R. Monnier, L. Degiorgi, and D. D. Koelin, Phys. Rev. Lett.
{\bf 56}, 2744 (1986)


\bibitem{int5} C. Huscroft, A. K. McMahan, and R. T. Scalettar, Phys. Rev. Lett.
{\bf 82}, 2342 (1999).

\bibitem{int5b} M. B. Z\"{o}lfl {\em et al.}
Phys. Rev. Lett. {\bf 87}, 276403 (2001).

\bibitem{int5a} K. Held, A. K. McMahan, and R. T. Scalettar, Phys. Rev. Lett.
{\bf 87}, 276404 (2001).

\bibitem{int4} K. Held {\em et al.}, Phys. Rev. Lett. {\bf 85}, 373 (2000).

\bibitem{int3} K. Held and R. Bulla, Eur. Phys. J. B. {\bf 17}, 7 (2000).

\bibitem{int6} P. van Dongen {\em et al.}, Phys. Rev. B. {\bf 64}, 195123
(2001).

\bibitem{int9} Y. Ono {\em et al.}, Eur. Phys. J. B. {\bf 19}, 375 (2001).

\bibitem{a1} U.Brandt and A.Giesekus, Phys. Rev. Lett. {\bf 68}, 2648 (1992).

\bibitem{a2} R. Strack, Phys. Rev. Lett. {\bf 70}, 833 (1993).

\bibitem{a3} I. Orlik and Z. Gul\'{a}csi, Phil. Mag. Lett. {\bf 78}, 177
(1998).

\bibitem{a4} I. Orlik and Z. Gul\'{a}csi, Phil. Mag. B. {\bf 81}, 1587 (2001);
Z. Gul\'{a}csi and I. Orlik, J. Phys. A: Math. Gen. {\bf 34},
L359, (2001).

\bibitem{negy} P. Gurin and Z. Gul\'{a}csi, Phys. Rev. B. {\bf 64}, 045118 (2001).

\bibitem{expl0} In $D=3$ the vector ${\bf r}$ can take 13 different values: $
{\bf x}_{\tau }$ ($\tau =1,2,3$), ${\bf x}_{\tau ^{\prime }}\pm
{\bf x}_{\tau }$ ($\tau ^{\prime }>\tau $), and ${\bf x}_{3}\pm
{\bf x}_{2}\pm {\bf x}_{1}$.

\bibitem{condi} The relations read, for example,
 $\sum_{n=1}^{8}|a_{n,f}|^{2}=K_{d}-E_{f}-U$,
 $\sum_{\beta _{1},\beta
_{2}=0,1}a_{n(0,\beta _{1},\beta _{2}),b}^{\ast }a_{n(1,\beta
_{1},\beta _{2}),b}=-t_{{\bf x}_{1}}^{b}$, $\sum_{\beta
_{1}=0,1}a_{n(0,0,\beta _{1}),b}^{\ast }a_{n(1,1,\beta
_{1}),b}=-t_{{\bf x}_{1}+{\bf x}_{2}}^{b}$, etc.

\bibitem{55} Here $D_{\beta ,\alpha }=(1-\delta
_{\beta ,-1})\delta _{\alpha ,0}$+$\delta _{\alpha ,\beta
}(1-\delta _{\alpha ,0})$, $n^{\pm }=n[g^{\pm }(\beta _{1},\alpha
_{1}),g^{\pm }(\beta _{2},\alpha _{2}),g^{\pm }(\beta _{3},\alpha
_{3})]$, with $g^{\pm }(\beta ,\alpha )=\beta \delta _{\alpha
,0}+(1\mp \alpha )(1-\delta _{\alpha ,0})/2$. The microscopic
parameters are contained in $T_{\bar {\bf r},\nu}^{b,b^{\prime
}}=-(1-\delta _{\nu ,0})[\delta _{b,b^{\prime }}t_{\bar {\bf
r}}^{b}+ (1-\delta _{b,b^{\prime }})V_{\bar {\bf r}}]-\delta _{\nu
,0}[\delta _{b,b^{\prime }}(\delta _{b,f}(E_{f}+U-K_{d})-\delta
_{b,d}K_{d})+(1-\delta _{b,b^{\prime }})\delta_{b,d}V_{0}]$, where
$\bar {\bf r}=\sum_{i=1}^{3}\alpha _{i}{\bf x}_{i}$ is a vector in
the unit cell $I$ (which thus determines the values of $\alpha
_{i}$ according to ref. \cite {expl0}), or $\bar {\bf r}=0$ (i.e.,
$\alpha _{i}=0$), and $\nu =\sum_{i=1}^{3}|\alpha _{i}|^{2}$.

\bibitem{condi3} Ground states with different total spin have been
discussed, e.g., by R. Arita and H. Aoki, Phys. Rev. B. {\bf 61},
12261 (2000).

\bibitem{cur}
Following D. Baeriswyl et al. [Phys. Rev. B. {\bf 35}, 8391
(1987)] the current operator is obtained for orthogonal ${\bf
x}_{\tau} $ as $\hat{\jmath}_{\tau }=(ie)/(\hbar
V)\sum_{b,b^{\prime }=d,f}\sum_{ {\bf j},\sigma ,{\bf r}}({T_{{\bf
r},\nu \neq 0}^{b,b^{\prime }}}^{\ast } \hat{b^{\prime }}_{{\bf
j}+{\bf r},\sigma }^{\dagger }\hat{b}_{{\bf j}\sigma }-T_{{\bf
r},\nu \neq 0}^{b,b^{\prime }}\hat{b}_{{\bf j}\sigma }^{\dagger }
\hat{b}_{{\bf j}+{\bf r},\sigma }^{\prime }){\bf r}\cdot {\bf
x}_{\tau }/| {\bf x}_{\tau }|.$

\bibitem{het} J. M. Wills, O. Eriksson, and A. M. Boring, Phys. Rev. Lett. {\bf
67}, 2215 (1991).

\bibitem{nyo} U. Hofmann and J. Keller, Z. Phys. B. {\bf 74}, 499 (1989).

\bibitem{kil} H.L. Schl\"{a}fer, G. Gliemann, {\em Einf\"{u}hrung in die
Ligandenfeldtheorie} (Akademische Verlagsgesellschaft, Frankfurt,
1980), p. 451, and Tables on p. 315-317.

\bibitem{sham} T. Portengen, Th. \"{O}streich, and L. J. Sham, Phys. Rev. B.
{\bf 54}, 17452 (1996).

\bibitem{exp} T. Ito {\em et al.}, Phys. Rev. B. {\bf 59}, 8923 (1999).

\bibitem{geb} E. H. Lieb and F. Y. Wu, Phys. Rev. Lett. {\bf 20}, 1445
(1968).
\end{references}
\end{document}